\documentclass[twocolumn,showpacs,preprintnumbers,amsmath,amssymb,superscriptaddress]{revtex4-1}
\usepackage{graphicx}
\usepackage{dcolumn}
\usepackage{bm}
\usepackage{float}
\usepackage{color}

\begin{document}

\selectfont

\title{\color{black}Suppressing Epidemics with a Limited Amount of Immunization Units}

\author{Christian M. Schneider}
\email{schnechr@ethz.ch}
\affiliation{Computational Physics, IfB, ETH Zurich, Schafmattstrasse 6, 8093 Zurich, Switzerland}
\author{Tamara Mihaljev}
\affiliation{Computational Physics, IfB, ETH Zurich, Schafmattstrasse 6, 8093 Zurich, Switzerland}
\author{Shlomo \surname{Havlin}}
\affiliation{Minerva Center and Department of Physics, Bar-Ilan University, 52900 Ramat-Gan, Israel}
\author{Hans J. Herrmann}
\affiliation{Computational Physics, IfB, ETH Zurich, Schafmattstrasse 6, 8093 Zurich, Switzerland}
\affiliation{Departamento de F\'{\i}sica, Universidade Federal do Cear\'a, 60451-970 Fortaleza, Cear\'a, Brazil}

\date{\today}
 
\begin{abstract}
{
The way diseases spread through schools, epidemics through countries, and viruses through the Internet is crucial in determining their risk. Although each of these threats has its own characteristics, its underlying network determines the spreading. To restrain the spreading, a widely used approach is the fragmentation of these networks through immunization, so that epidemics cannot spread. Here we develop an immunization approach based on optimizing the susceptible size, which outperforms the best known strategy based on immunizing the highest-betweenness links or nodes. We find that the network's vulnerability can be significantly reduced, demonstrating this on three different real networks: the global flight network, a school friendship network, and the internet. In all cases, we find that not only is the average infection probability significantly suppressed, but also for the most relevant case of a small and limited number of immunization units the infection probability can be reduced by up to $55\%$.}
\end{abstract}



\maketitle
\section{Introduction}
\begin{figure*}\hspace*{-.5cm}
  \includegraphics[width=4.75cm,angle = -90]{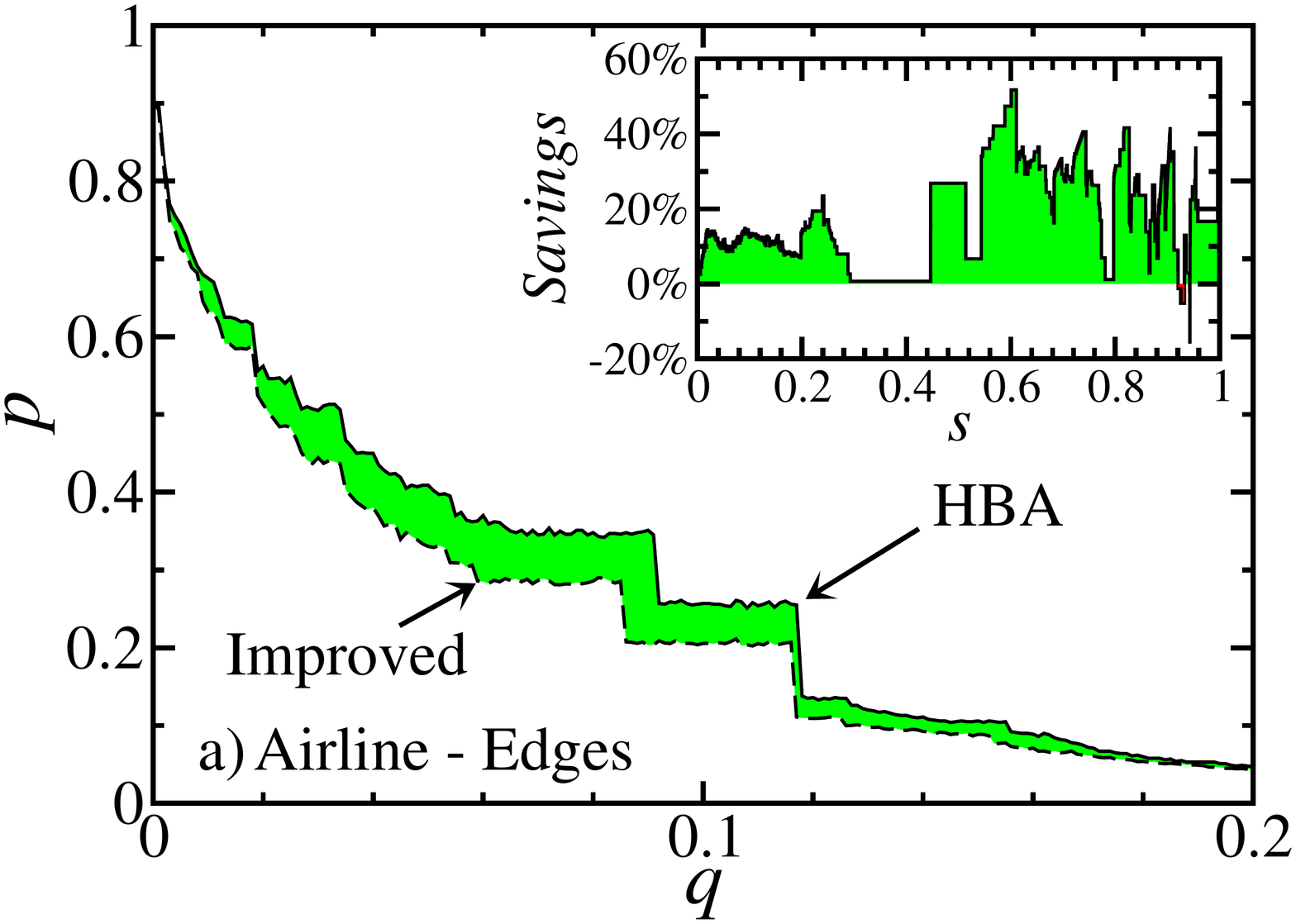}\hspace*{-.5cm}
  \includegraphics[width=4.75cm,angle = -90]{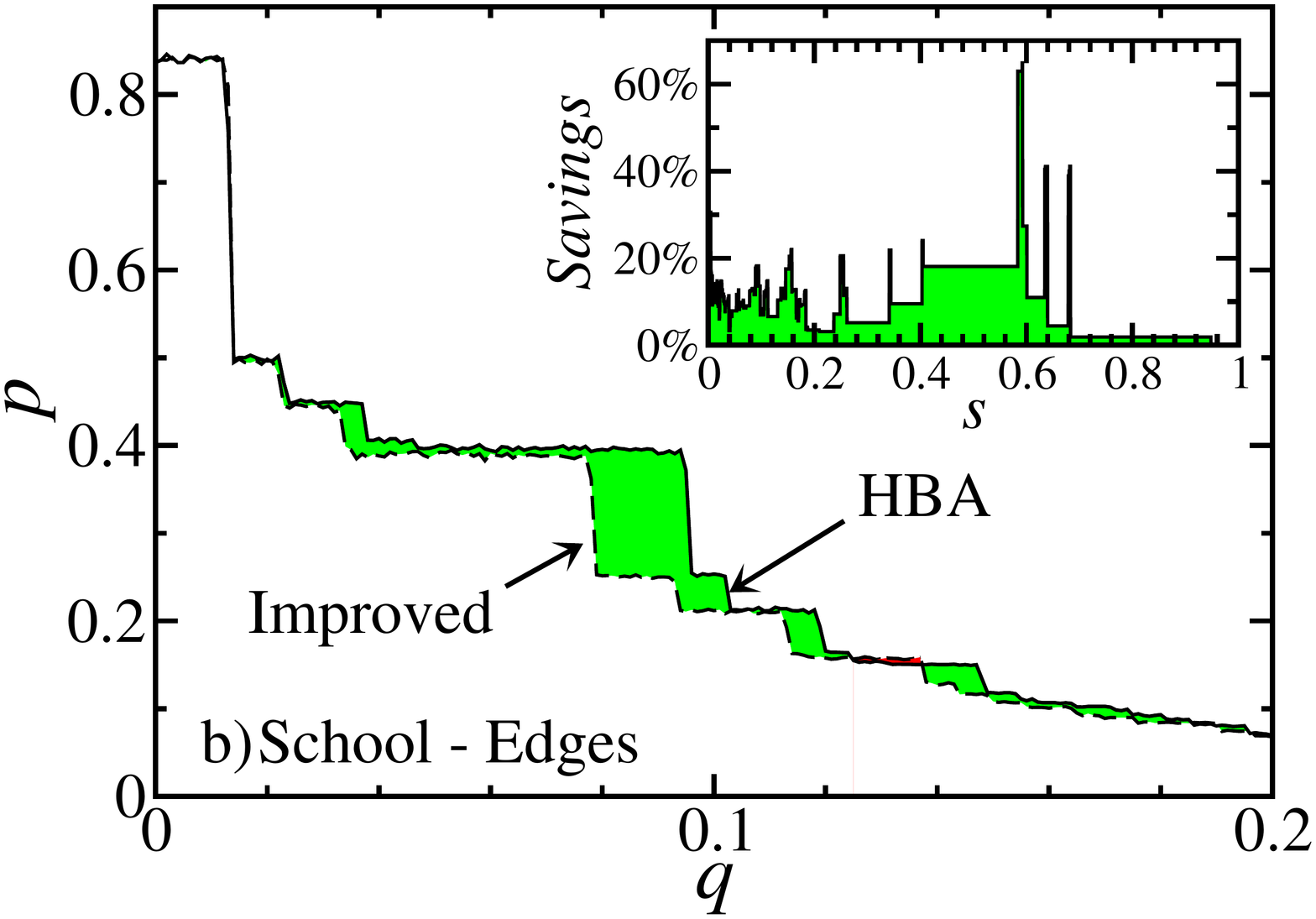}\hspace*{-.5cm}
  \includegraphics[width=4.75cm,angle = -90]{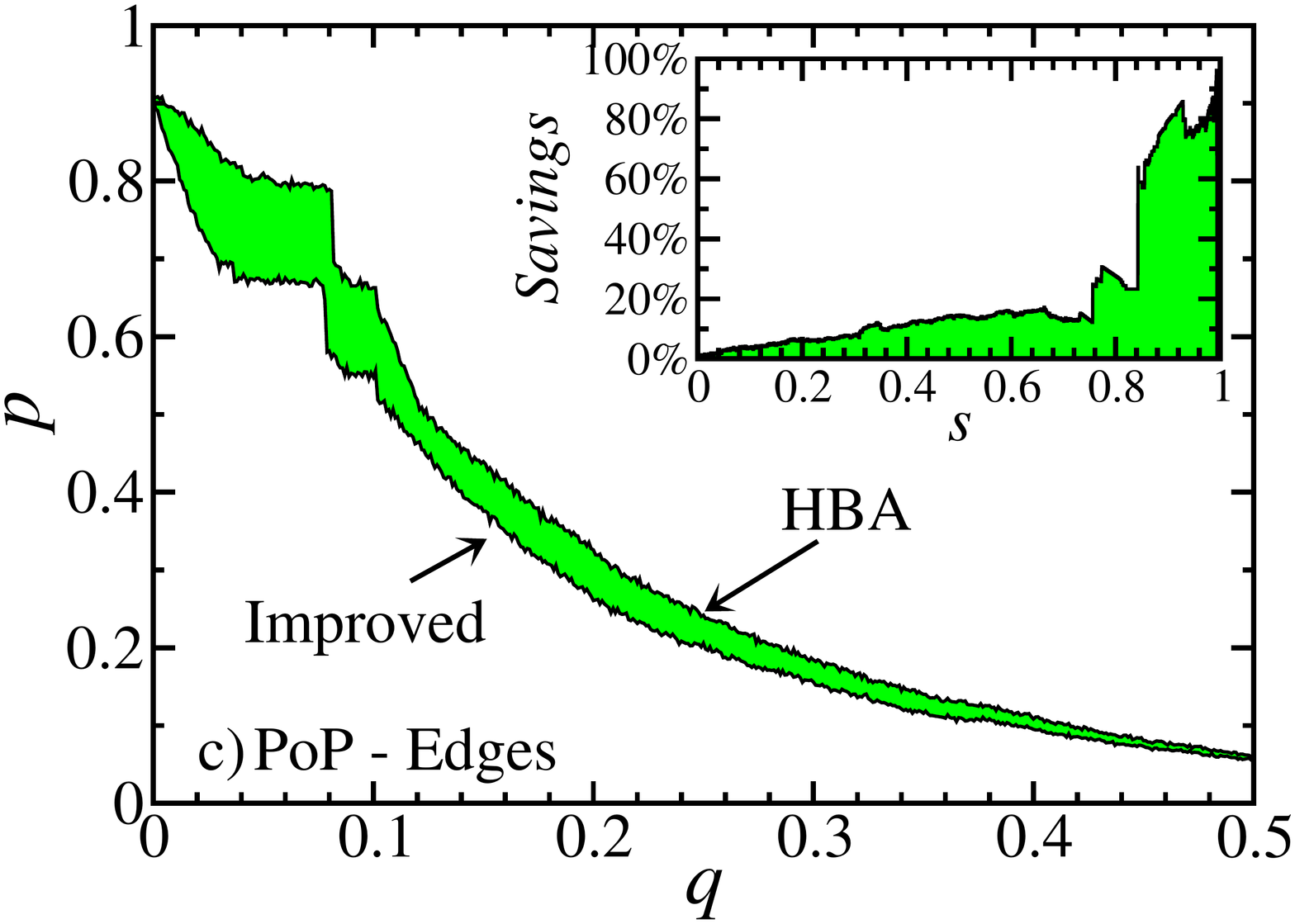}
  \caption{\color{black}(Color online) Comparison between the two edge immunization strategies, betweenness based (full line) and our improved immunization (dashed line). Plotted is the probability of becoming infected, $p$, for (a) the global airline network ($N = 3666$ and $M = 27 235$), (b) a school friendship network ($N = 1461$ and $M = 3875$), and (c) the PoP internet network ($N = 1098$ and $M = 6089$) as a function of the immunized edge fraction $q$. Both betweenness-based and improved immunization can reduce the spreading of diseases significantly. Nevertheless, for practical cases of small fractions of immunized edges, the improved strategy is significantly more efficient. In the insets the savings of immunization units obtained from our immunization strategy are shown. We compare the number of immunization units required to keep the fraction of affected nodes under a certain value and plot the relative number of saved units vs the maximal size of the spreading, $s$. The average savings are $17.8\%$, $9\%$, and $21.2\%$ for the airline, school, and internet networks, respectively.}
\label{fig:4}
\end{figure*}

Every few years a potential global pandemic like severe acute respiratory syndrome (SARS) or swine flu occurs \cite{Colizza07b}. Nowadays there is a possibility to reach every city in the world within at most a day, which allows for the evolution of a local disease to a global pandemic \cite{Anderson92,Liljeros01,Lloyd01,Pastor01,Barthelemy04,Colizza07a,Colizza07c}. Crucial for fast global disease spreading is the international flight network \cite{Colizza06,Bobashev08}. To avoid spreading through this network, significant effort is made by, e.g., screening passengers or canceling flights. Since it is unrealistic to examine all passengers, it is vital to apply the best possible immunization strategy in order to most effectively exploit limited resources, like available vaccines and man power.\\
{Different strategies to immunize nodes or links of a network have been studied in the past \cite{Albert00,Callaway00,Cohen01,Pastor02,Keeling05,Christakis10}. Targeted immunization strategies, in which nodes or links playing a special role in the network architecture are immunized first, have proven to be very effective \cite{Holme02,Cohen03,Holme04,Chen08}. The strategy believed to be the most efficient is the targeted immunization based on immunizing the highest betweenness centrality nodes or links \cite{Holme02}. The adaptive betweenness centrality is the number of shortest paths passing through the node or the link, recalculated for the network of non immunized nodes at each step of the immunization process.\\
Here we show that in fact this strategy is not optimal and a significantly more efficient immunization strategy exists. This is important since improving the effectiveness of immunization strategies even by a small amount can result in saving thousands of human lives.}\\
Based on percolation we developed a more efficient immunization strategy using the betweenness centrality measure \cite{Holme02}. We studied quantitatively the effectiveness of the improved strategy on real networks, the global airport network \cite{Li11}, the friendship network in American schools \cite{Gonzalez06}, and the internet at the level of service providers (point of presence, PoP) \cite{PoP}, as well as on model networks. We study both cases of immunization, the immunization of nodes, which models, for instance, immunization of people or airports, and the immunization of links, modeling, for instance, prevention of direct contact between people or the performance of special checks and vaccinations on all flights between two specific airports. We find that, although very effective, the immunization strategy based on immunizing nodes or edges with the highest values of dynamically recalculated betweenness is not optimal. Starting from such a strategy, our approach reduces the risk of becoming infected on average by more than $10\%$, independent of the amount of immunization units for the real examples of airport, friendship, and internet networks. Moreover, for specific numbers of immunization units the improvement is even up to $55\%$. For model networks, the improvement is close to $30\%$ on average, and a maximal improvement of over $80\%$ can be obtained or up to $29\%$ of immunization units can be saved.\\
\begin{figure*}
 \includegraphics[width=1.cm,angle = 0]{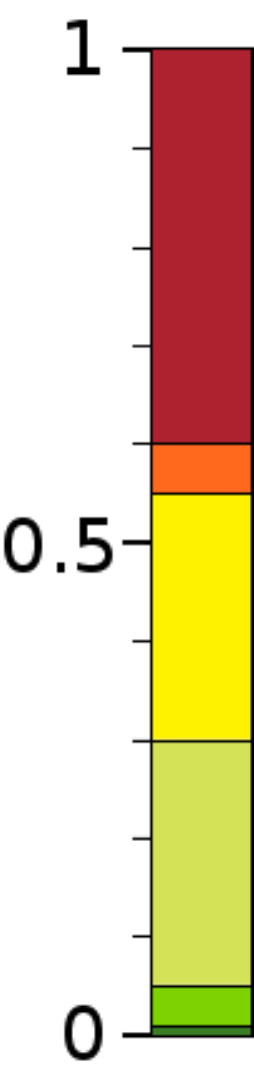}\hspace*{0.2cm}
 \includegraphics[width=5.3cm,angle = 0]{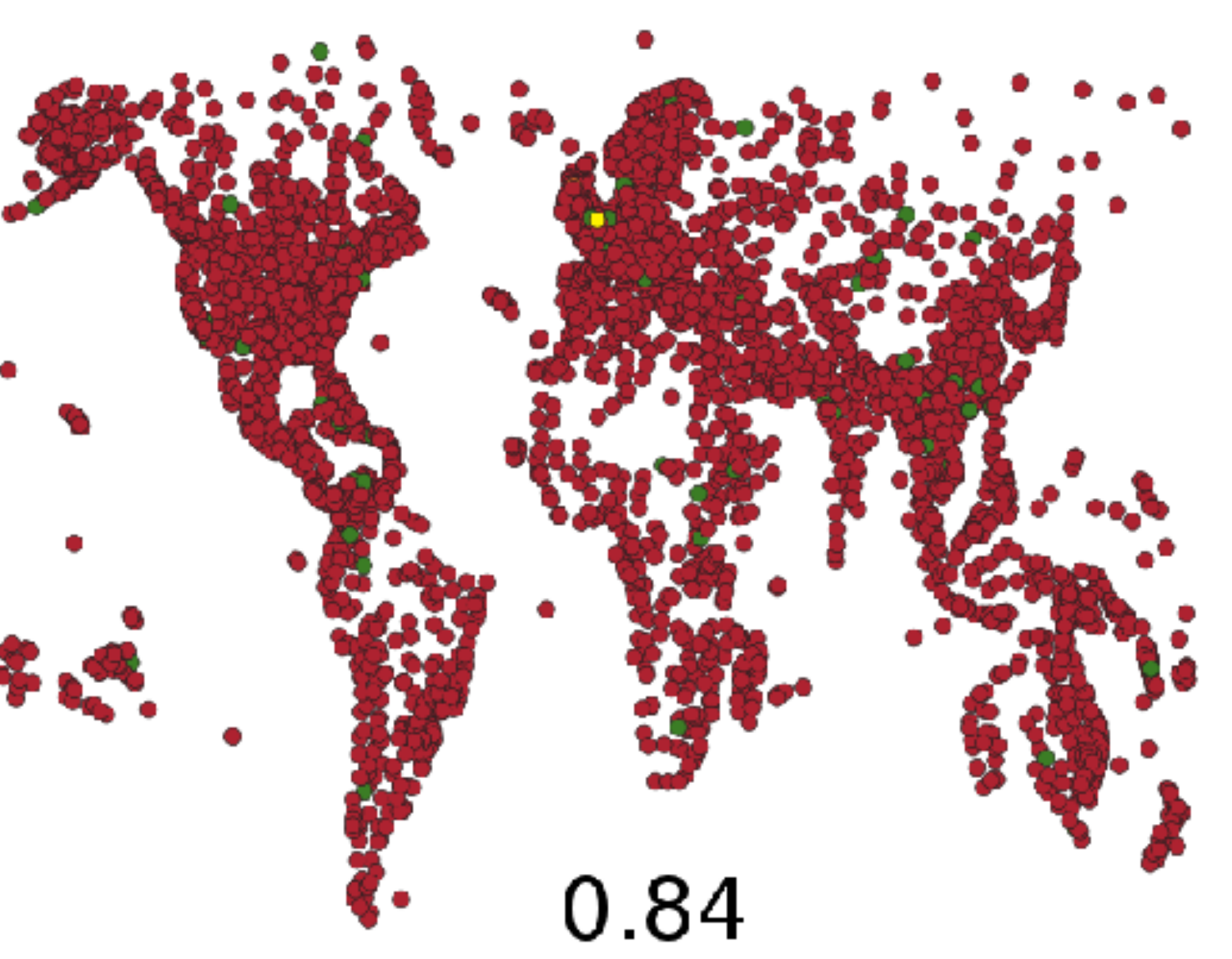}\hspace*{0.2cm}
 \includegraphics[width=5.3cm,angle = 0]{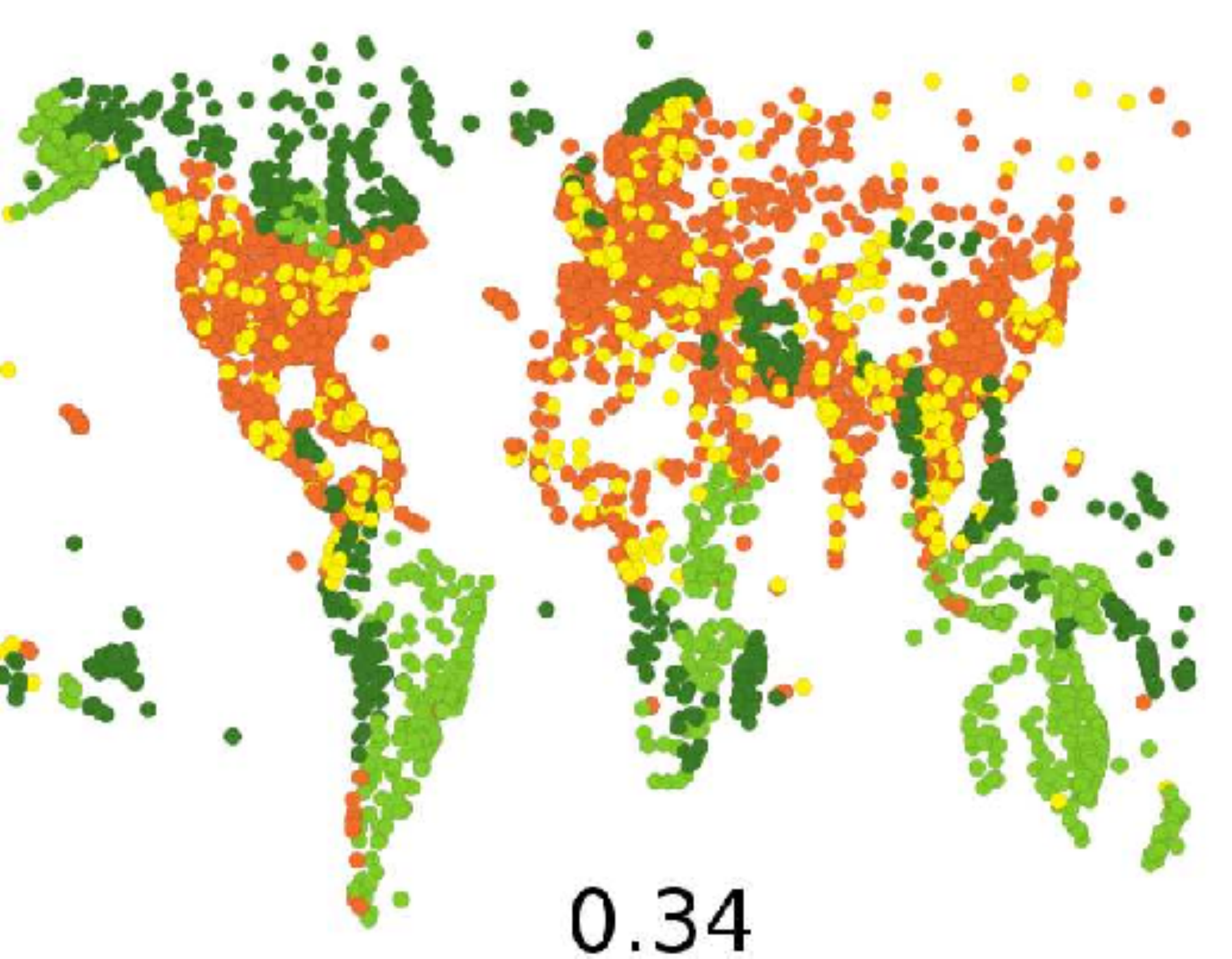}\hspace*{0.2cm}
 \includegraphics[width=5.3cm,angle = 0]{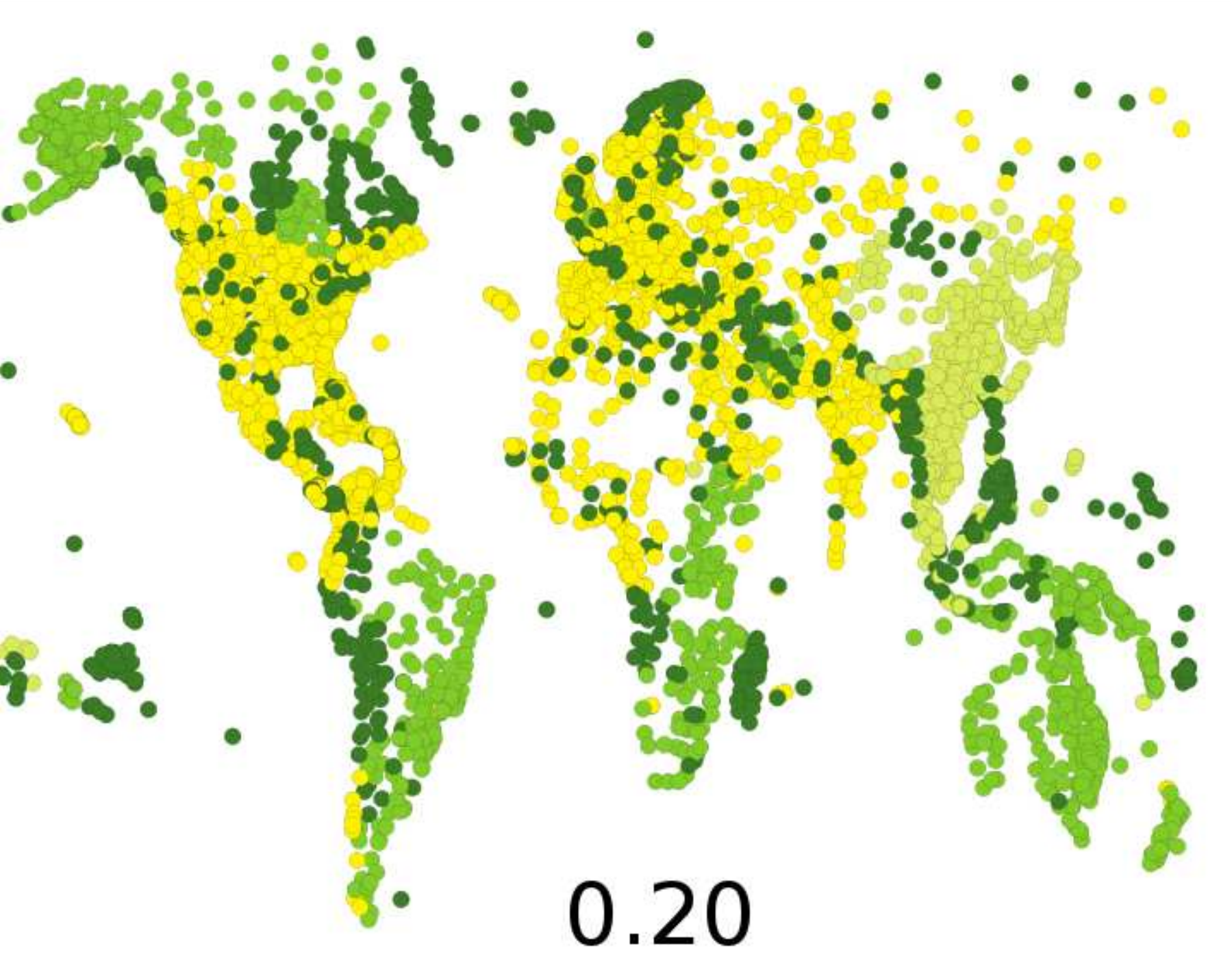}
 \caption{(Color online) The simulation of an epidemic using the SIR model with $\alpha = 0.2$ and $\beta = 0.05$ for the global airline network, starting from a single randomly infected node. To suppress the spreading of the disease, a small number of $q = 0.09$ flights are immunized according to a random (left), betweenness (middle), and improved (right) strategy. The color code represents the probability of becoming infected. Note that the average probability of becoming infected is $84\%$ for random, $34\%$ for betweenness, and $20\%$ for our improved immunization method. The reduction of the infection probability is due to the efficient decoupling of the regions; thus passengers are screened on their journey between these regions, e.g., from China to Europe.} The picture is created with pajek \cite{Pajek08}.
\label{fig:airportsDots}
\end{figure*}
The effectiveness of our approach is illustrated in Fig. \ref{fig:airportsDots}, where we compare the disease spreading (based on the susceptible-infected-recovered (SIR) model \cite{Anderson92}) in the global airport network immunized at random (left), using the high betweenness based link immunization strategy (middle) \cite{Holme02}, which is already significantly better, and finally using our developed strategy (right), which is more than $40\%$ more effective than the recalculated high-betweenness method. In all cases the same number of links (flights) are immunized. The color code represents the probability that the node becomes infected, going from green (dark gray, right) for low probability over yellow (light gray, right) and orange (gray, middle) to red (dark gray, left) for very high probability of infection. While for the random case many immunization units are wasted, the betweenness based strategy suppresses the spreading from certain regions to the backbone of the network. Nevertheless, our improved strategy is more efficient in identifying these regions and thus more efficient in suppressing diseases. For example the two large jumps at $9\%$ and $12\%$ of immunized edges in Fig. \ref{fig:4}(a) correspond to the decoupling of East Asia from Central America, Europe, and the United States (2336 vs 2503 immunized edges) and the decoupling of Europe from America (3169 vs 3192 immunized edges), respectively.
\\
\section{Methods}
We use the susceptible size of the network that could be infected as the performance measure of the immunization procedure. A good immunization strategy should make such a susceptible size as small as possible. The susceptible size $R$ is defined here as the sum of the sizes of the largest connected clusters $S(q)$ of the networks of non immunized nodes remaining after immunization of $q$ nodes or edges:
\begin{eqnarray}
R = \frac{1}{(N + 1)N} \sum_{q = 0}^{N} S(q),
\end{eqnarray}
where $N$ is the size of the network \cite{Schneider10b}. This measure captures the network response to immunization throughout the whole immunization process, and not only on the percolation threshold at which the network of non immunized nodes becomes disconnected.\\
In our search for a more efficient immunization strategy of nodes or edges, we start from the efficient known strategy of high betweenness. First, we calculate the sequence in which nodes or edges would be immunized if the high-betweenness adaptive (HBA) immunization strategy \cite{Holme02} was used. To this end, we calculate the shortest path between all pairs of nodes and count how often a node or edge lies on a shortest path. This number defines its betweenness centrality. The node or edge with the highest contribution to the shortest paths is identified and immunized. Repeating this procedure with the remaining network of non immunized nodes until it vanishes leads to the high-betweenness immunization sequence.\\
\begin{figure}
  \includegraphics[width=8.5cm,angle = 0]{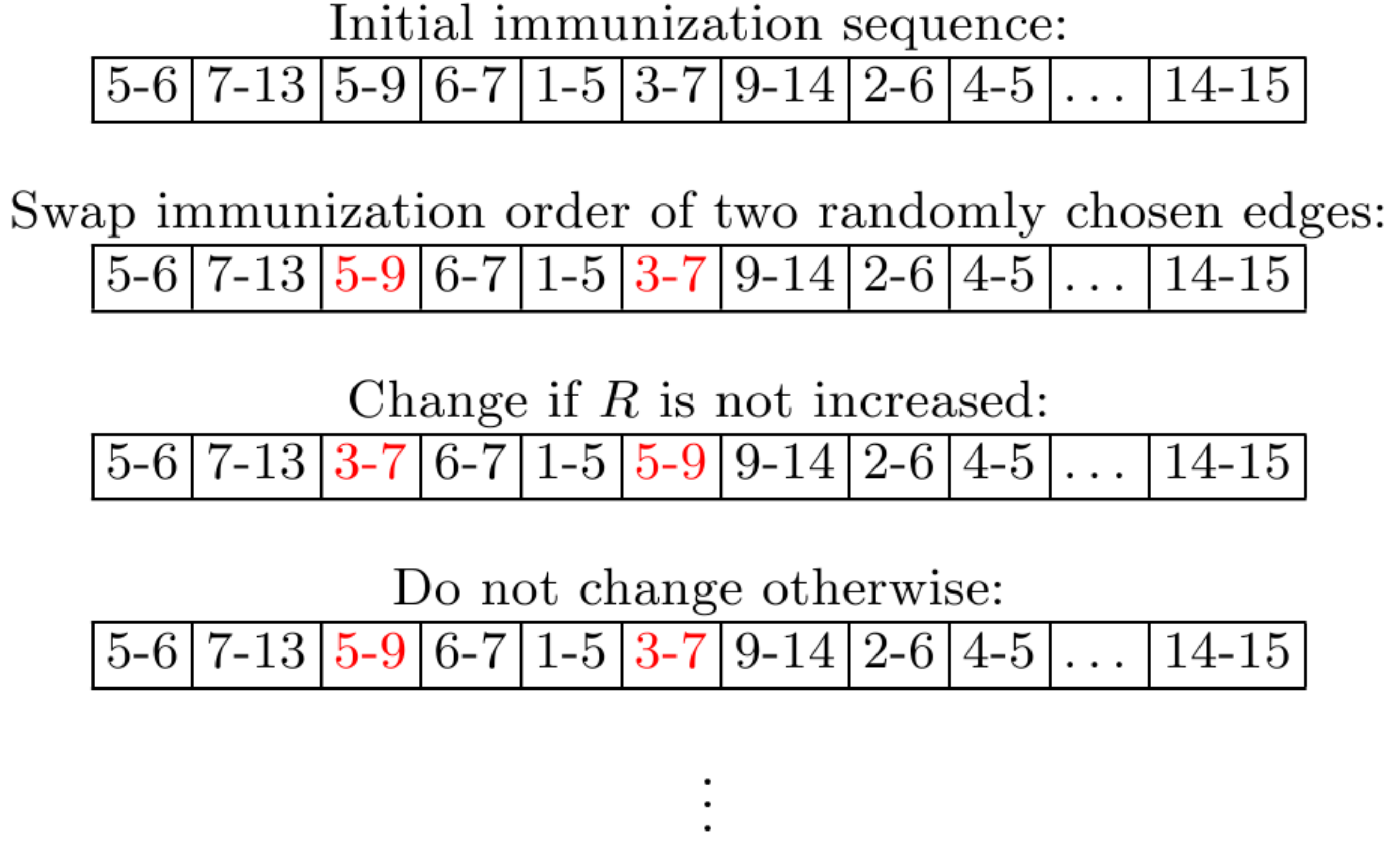}
  \caption{(Color online) Illustration of improving the edge immunization strategy. Starting from an initial immunization sequence (based on betweenness centrality), two randomly chosen edges (in this case the edge between nodes 5 and 9 and the edge between 3 and 7) are swapped in the immunization sequence. If the swap does not increase the susceptible size $R$ of the network, it is kept; otherwise the swap is withdrawn.}
\label{fig:1}
\end{figure}
To improve this immunization strategy, we modify the initial, high-betweenness-based, immunization sequence using the following algorithm. We choose two nodes or edges randomly, switch the order in which they are immunized, and recalculate the susceptible size $R$, Eq. (1), according to the new sequence. If the susceptible size does not increase when the new immunization sequence is used, the change in the sequence is accepted. The sequence is further improved by repeating the same procedure many times.\\
To further improve our strategy, we use in addition an improving algorithm employing a population of immunization sequences. We start from $1000$ equal high-betweenness immunization sequences and choose one of them randomly to perform a swap. If the susceptible size of the network after immunizing with the new sequence is not increased, it replaces the sequence from the population for which the network has the highest susceptible size. The final immunization sequence is the one with the lowest susceptible size. The basic algorithm for improving edge immunization sequences is demonstrated in Fig. \ref{fig:1}.\\
Note that the algorithm uses the global knowledge of the network to achieve the final immunization sequence. The calculation of the initial sequence, which is based on the betweenness centrality, and the optimization strategy are computationally costly; thus our algorithm scales usually much more slowly than linearly with system size. Nevertheless, we are able to obtain immunization sequences for networks with up to $N = 8000$ nodes and $M = 16 000$ edges in a reasonable time.
\section{Results: Real networks}
\begin{figure*}\hspace*{-.5cm}
  \includegraphics[height=4.75cm,angle = 0]{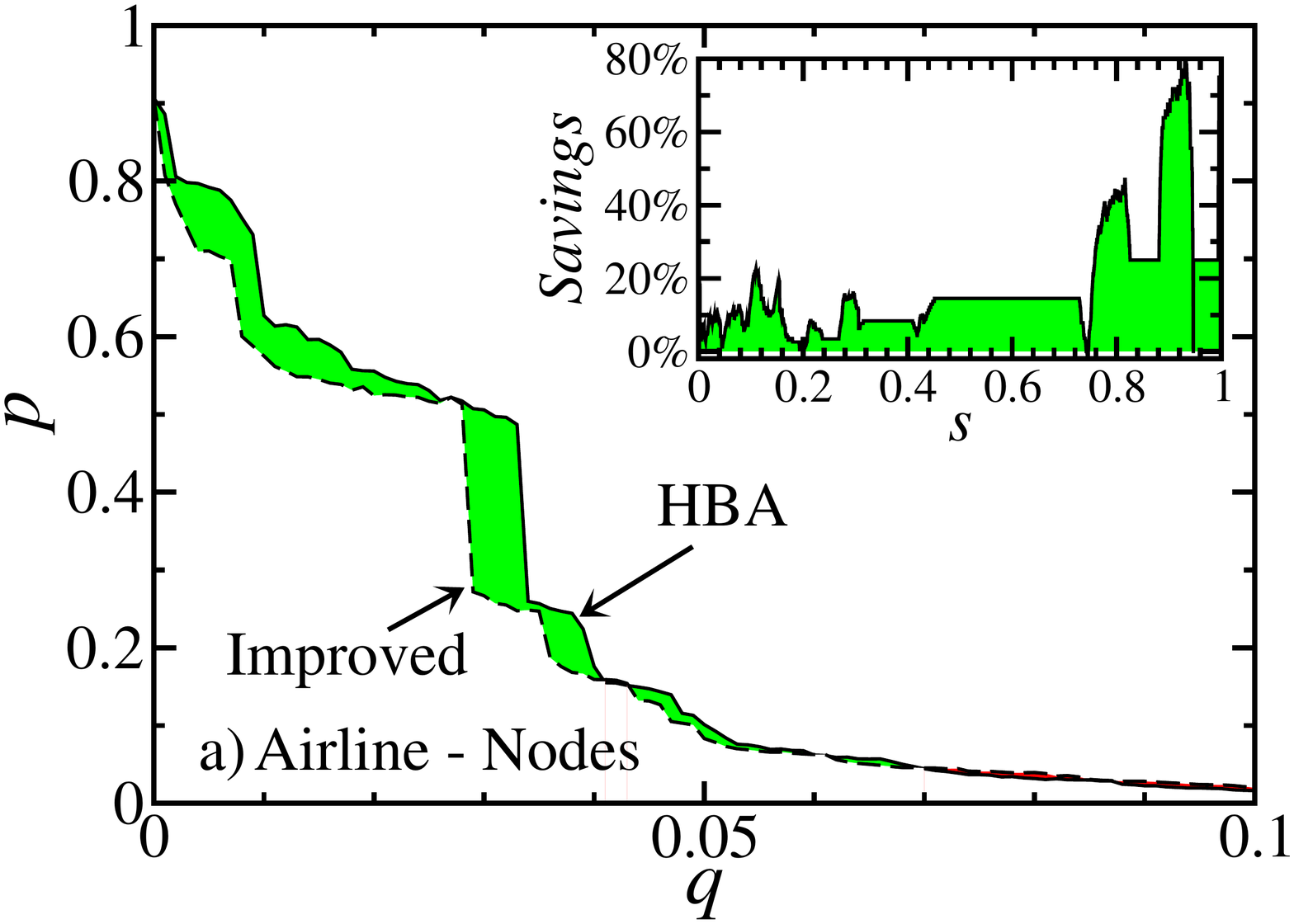}\hspace*{-.5cm}
  \includegraphics[height=4.75cm,angle = 0]{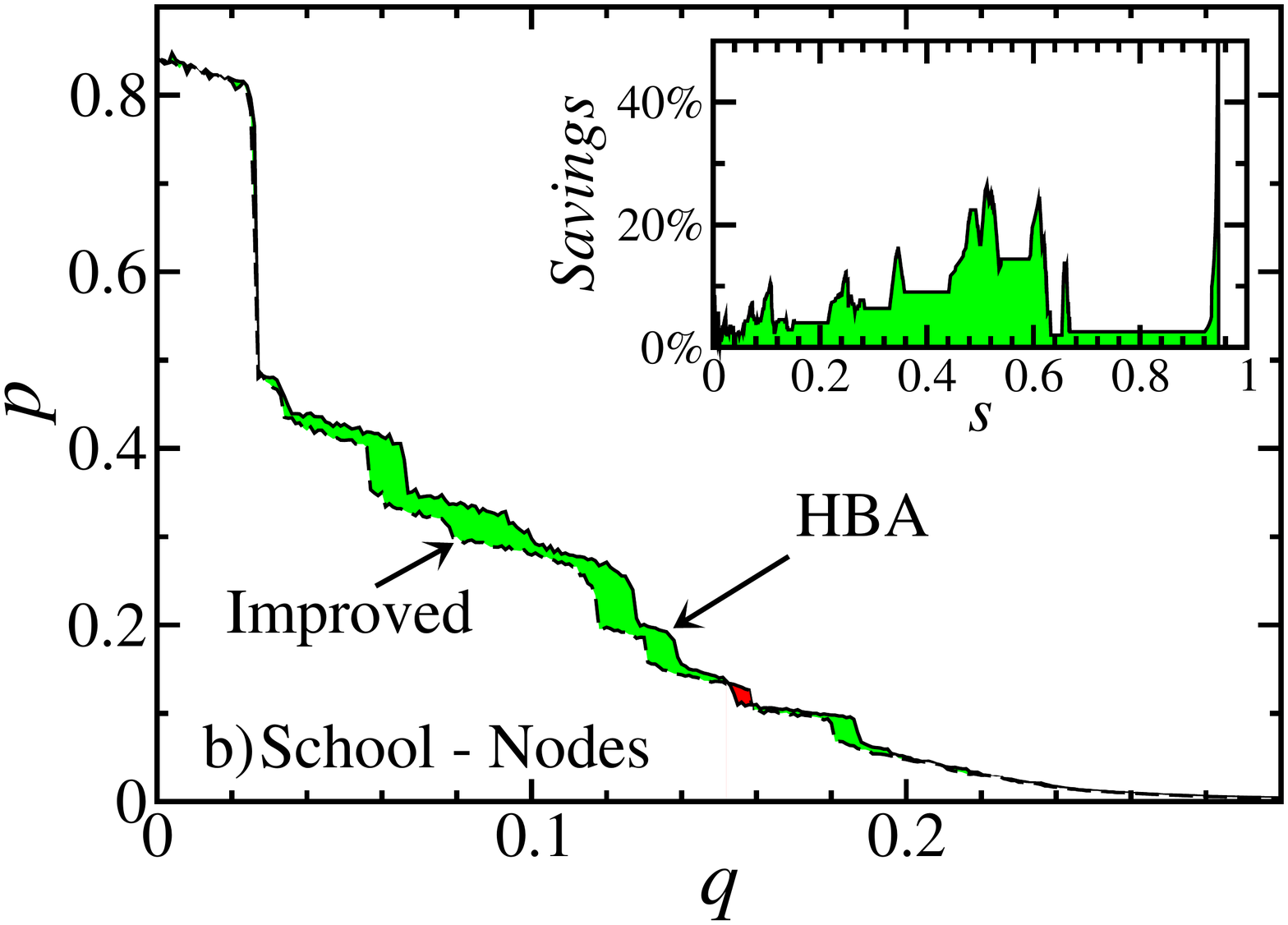}\hspace*{-.5cm}
  \includegraphics[height=4.75cm,angle = 0]{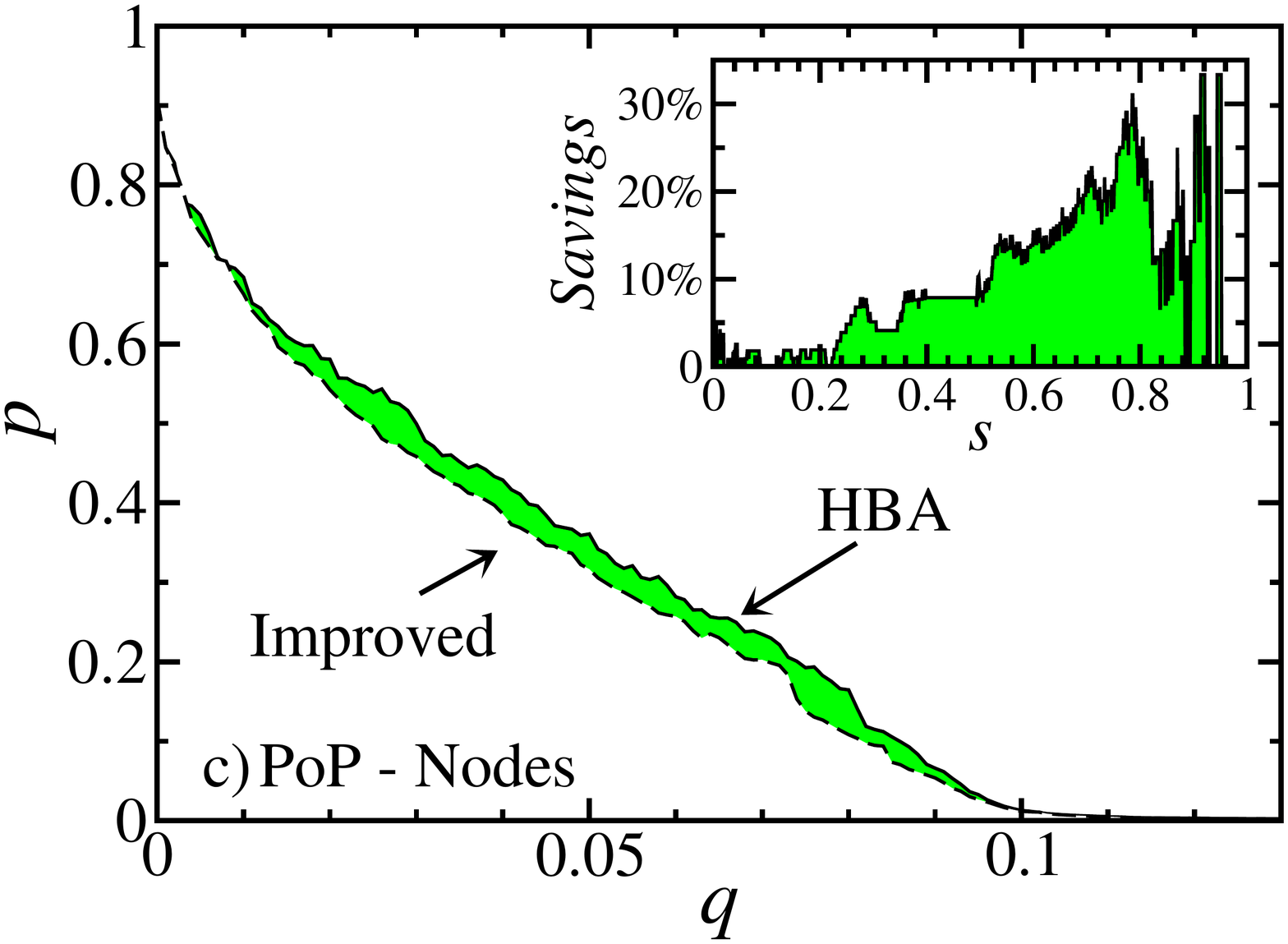}
  \caption{(Color online) Comparison between the two node immunization strategies, betweenness based (full line) and our improved immunization (dashed line). Plotted is the probability of becoming infected, $p$, for (a) the global airline network ($N = 3666$ and $M = 27 235$), (b) a school friendship network ($N = 1461$ and $M = 3875$), and (c) the PoP internet network ($N = 1098$ and $M = 6089$) as a function of the immunized node fraction $q$. Both betweenness-based and improved immunization can reduce the spreading of diseases significantly. Nevertheless, for practical cases of small fractions of immunized nodes, the improved strategy is significantly more efficient. {\color{black} In the insets the savings of immunization units obtained using our immunization strategy are shown. We compare the number of immunization units required to keep the fraction of affected nodes under a certain value and plot the relative number of saved units vs the maximal size of the spreading, $s$. The average savings are $18\%$, $7.2\%$, and $9.6\%$ for the airline, school, and internet networks, respectively.}}
\label{fig:Node}
\end{figure*}

To study our improved immunization strategy, we analyze its efficiency on two real network examples through which epidemics are spread, namely, the global airline network \cite{Li11,Colizza06} and school friendship networks \cite{Gonzalez06}. We also analyze the case of the internet at the level of point-of-presence as well as two models, the Erd\H{o}s-R\'enyi networks \cite{Erdos} and scale-free networks \cite{Barabasi} generated using the configurational model \cite{static}. For a typical friendship network we show the results for a single school, since we found that the other $81$ school networks we studied all behave in a qualitatively similar way. For all real networks we demonstrate the efficiency of our method for both link (Fig. \ref{fig:4}) and node immunization (Fig. \ref{fig:Node}).\\
For the airline network, an infected airport implies that sick people arrive at or depart from it. Consequently immunization means identifying sick people and inhibiting their travel. For example, link immunization can be done by screening people on specific flights - as done in several countries (including Japan and China) during the SARS and swine flu epidemics - while in the case of node immunization all people at an airport are screened. A more drastic immunization would be the canceling of flights or even the shut down of entire airports. For the school network, link immunization corresponds to preventing direct contacts between students and node immunization can be performed by immunizing students or temporarily removing them from classes. For the internet, the screening of traffic between routers is equivalent to link immunization and the upgrade of the router soft- or hardware to node immunization.\\
The efficiency of our method for improving immunization is evaluated by three different measures: modeling epidemic spreading using the SIR model, the saved immunization doses, and the susceptible size, which represents the upper limit of the number of infected individuals.\\
First we analyze the probability for a single node (airport, student, router) to become infected. To study this probability we simulate a disease spreading model (SIR) on the global airline transportation network with $N = 3666$ airports and $M = 27 235$ flight connections, a typical friendship network with $N = 1461$ students and $M = 3875$ friendship connections, and the internet with $N = 1098$ and $M = 6089$ connections. We simulate the SIR model \cite{Anderson92,Pastor01} with the parameters  $\alpha = 0.2$ (probability that the infection spreads from an infected node to its neighbor in one time step) and $\beta = 0.05$ (probability that a node recovers in a unit time step) starting from a single random infected node and averaging over $10 000$ independent runs.\\
\begin{figure*}
\includegraphics[height=4.75cm,angle = 0]{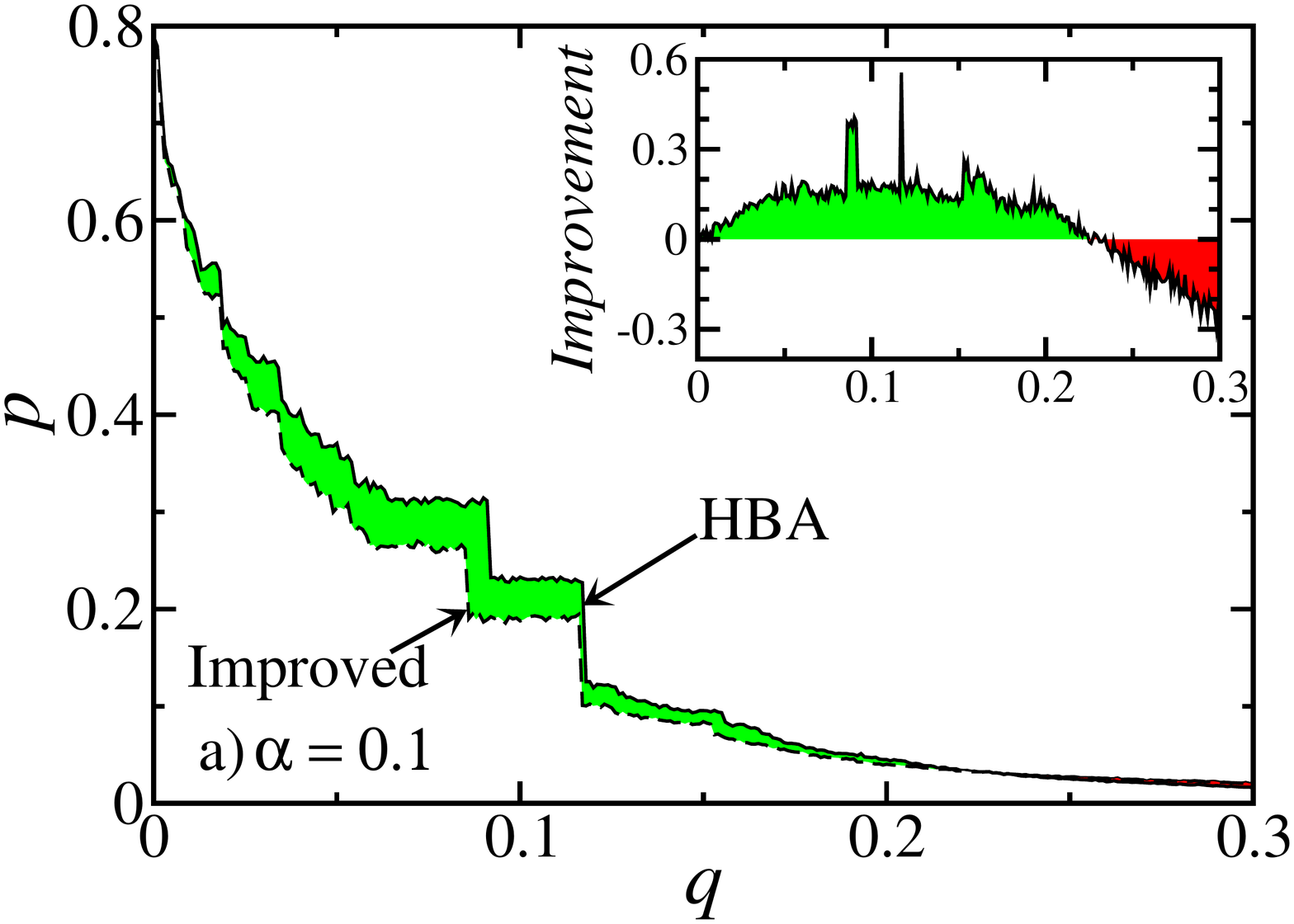}\hspace*{-.5cm}
\includegraphics[height=4.75cm,angle = 0]{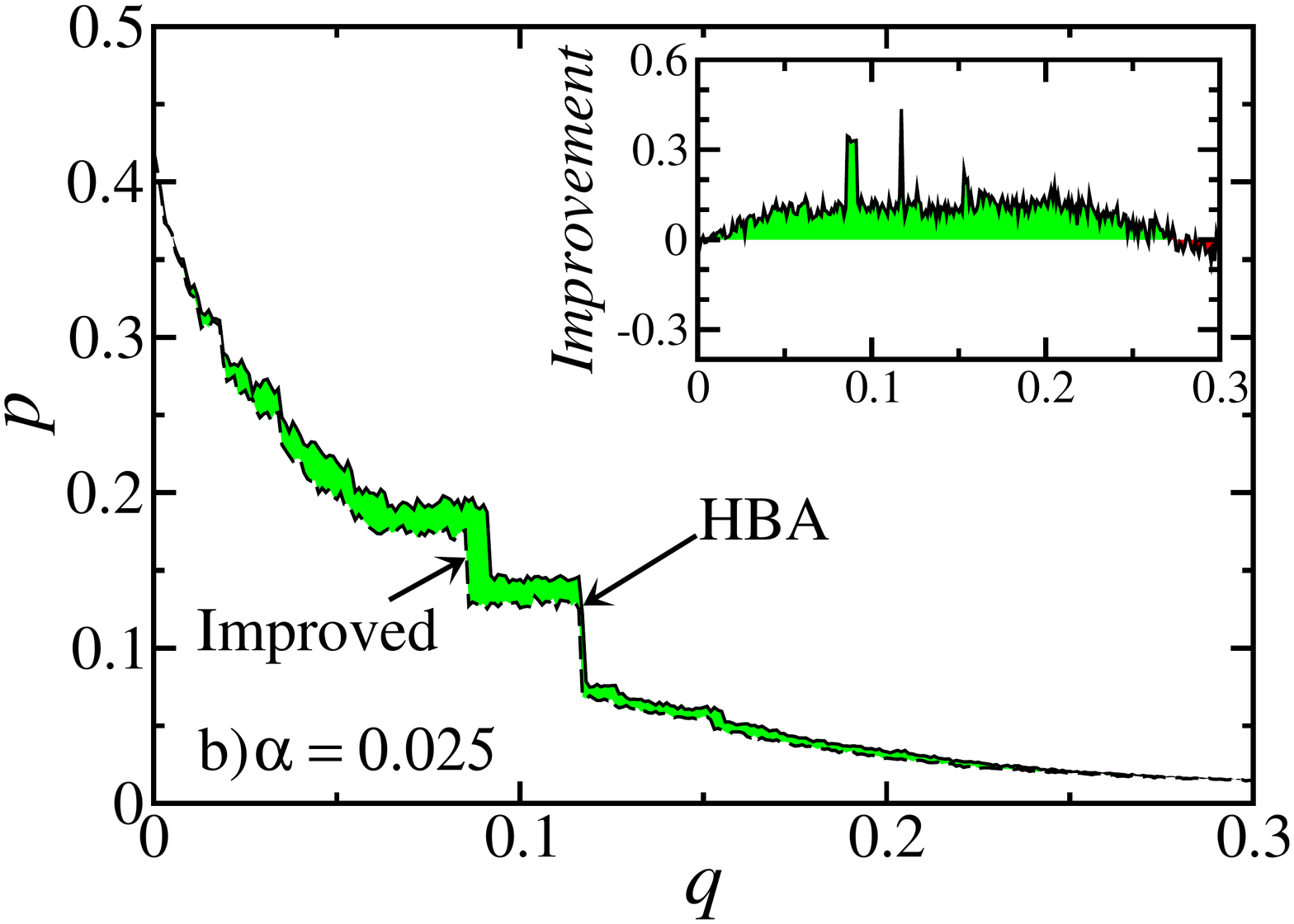}\hspace*{-.5cm}
\includegraphics[height=4.75cm,angle = 0]{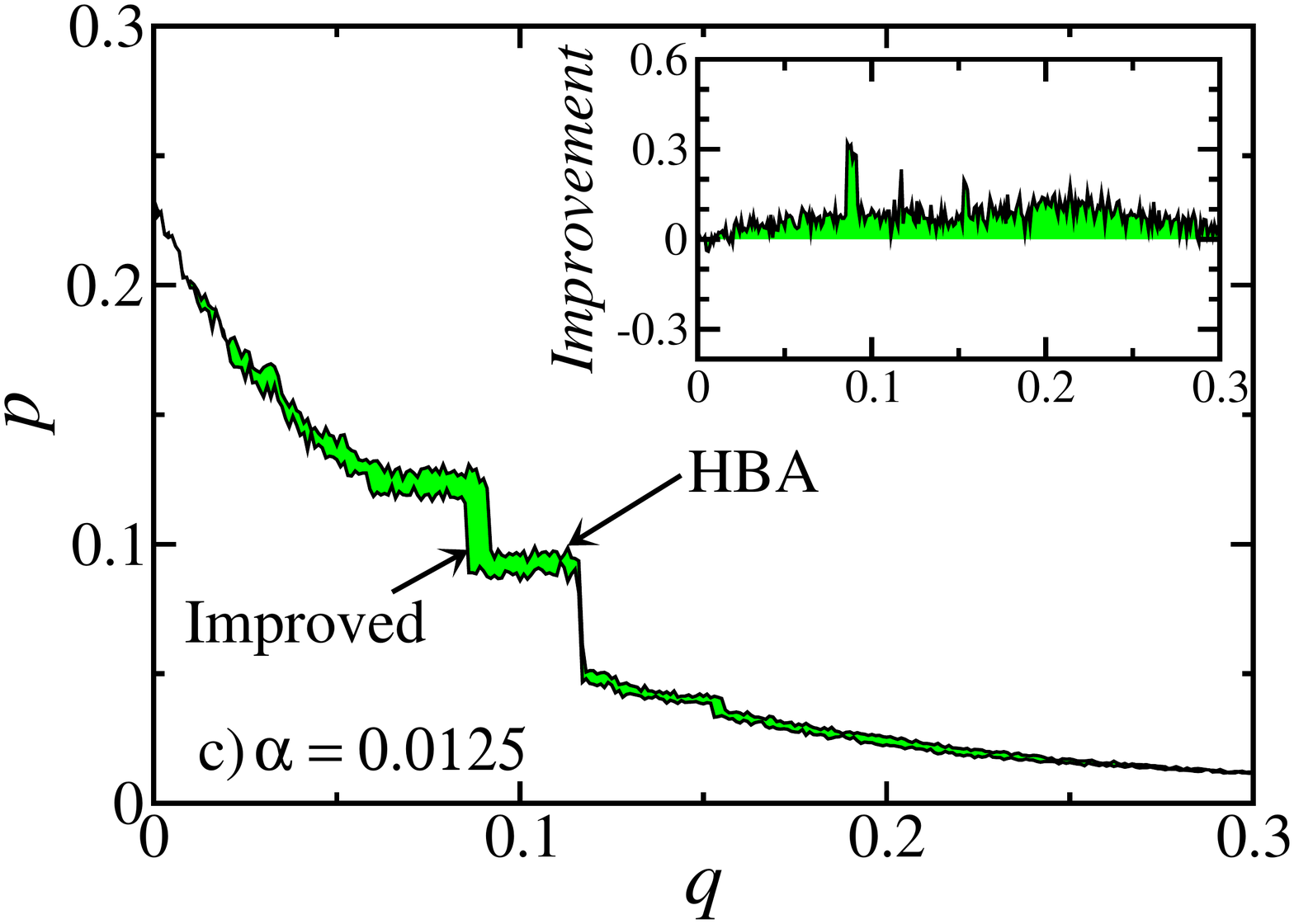}
\caption{(Color online) Comparison between the two immunization strategies, betweenness based (full line), and our improved immunization (dashed line), for different infection parameters $\alpha$. Plotted is the probability of becoming infected, $p$, for link immunization in the global airline network as a function of the immunized fraction $q$ for (a) $\alpha = 0.1$, (b) $\alpha = 0.025$, and (c) $\alpha = 0.0125$. {\color{black} For the practical cases of low fractions of immunized edges or nodes, our improved strategy is more efficient for all studied $\alpha$ values.} In the insets the relative improvement of our method is shown.}
\label{fig:less}
\end{figure*}
In Fig. \ref{fig:4}(a) we show the dependence of the infection probability on the fraction of immunized flights for airports. The full line is the probability of becoming infected in the largest connected cluster after immunizing or removing a fraction $q$ of edges, according to the high-betweenness immunization strategy. The dashed line corresponds to our improved immunization strategy. Thus, the green (light gray) area represents the improvement. For the most practical case of a small number of immunized flights, the improved immunization strategy is significantly more efficient, while for large $q$ the two strategies have nearly similar effects. {\color{black}Figure \ref{fig:4}(b) shows the probability that a student becomes infected as a function of the percentage of suppressed contacts, and Fig. \ref{fig:4}(c) shows the dependency of the probability that a router becomes infected on the percentage of controlled connections. Qualitatively, in all cases a similar improvement is observed for improved immunization. For the airline network, the average improvement is about $15\%$ for immunization fractions less than $20\%$, while the maximal improvement is about $55\%$ for $q \approx 11.9\%$. For the school network an average improvement of about $7\%$ for immunization fractions less than $20\%$ is obtained with a maximal improvement of $45\%$ for $q \approx 9.4\%$. In the case of the internet the average improvement is $15\%$ for fewer than $20\%$ controlled connections and the maximal improvement is $24\%$ for $q \approx 8\%$. Note that the simulations are for a highly contagious illness or virus. Results for less contagious ones are shown for the airline network in Fig. \ref{fig:less}.}\\
Not only is the probability for a single person to become infected of interest, but also the maximal possible spreading size is crucial. Therefore, we analyze the change of the largest connected cluster of non immunized nodes during the immunization process, a measure of the network's susceptible size. The size of this cluster gives the upper limit of the number of people that could become infected if the disease spreading starts on a node in this cluster. An efficient immunization strategy should keep this number as small as possible in all steps of the immunization process.\\
In the insets of Fig. \ref{fig:4} we show the possible savings of immunization units with our strategy. To reduce the maximal spreading size to a certain value $s$, our strategy needs significantly fewer units than the betweenness-based strategy . On average the potential savings are $9\%$ (school), $17.9\%$ (airline), and $21.2\%$ (PoP).\\
Our strategy is more efficient not only for edge immunization, but also for immunization of nodes. The results for node immunization for the same three networks are shown in Fig. \ref{fig:Node}. For the airline network, the average improvement is about $11\%$ for immunization fractions less than $7\%$, while the maximal improvement is about $49\%$ for $q \approx 3.3\%$. For the school network, an average improvement of about $8\%$ for immunization fractions less than $20\%$ is obtained with a maximal improvement of $34\%$ for $q \approx 18.5\%$. In the case of the internet the average improvement is $12\%$ for fewer than $10\%$ controlled connections and the maximal improvement is $34\%$ for $q \approx 8\%$. The average possible savings are $18\%$, $7.2\%$, and $9.6\%$ for the airline, school, and internet networks, respectively.

\section{Results: Model networks}
\begin{figure*}
  \includegraphics[width=4.75cm,angle = -90]{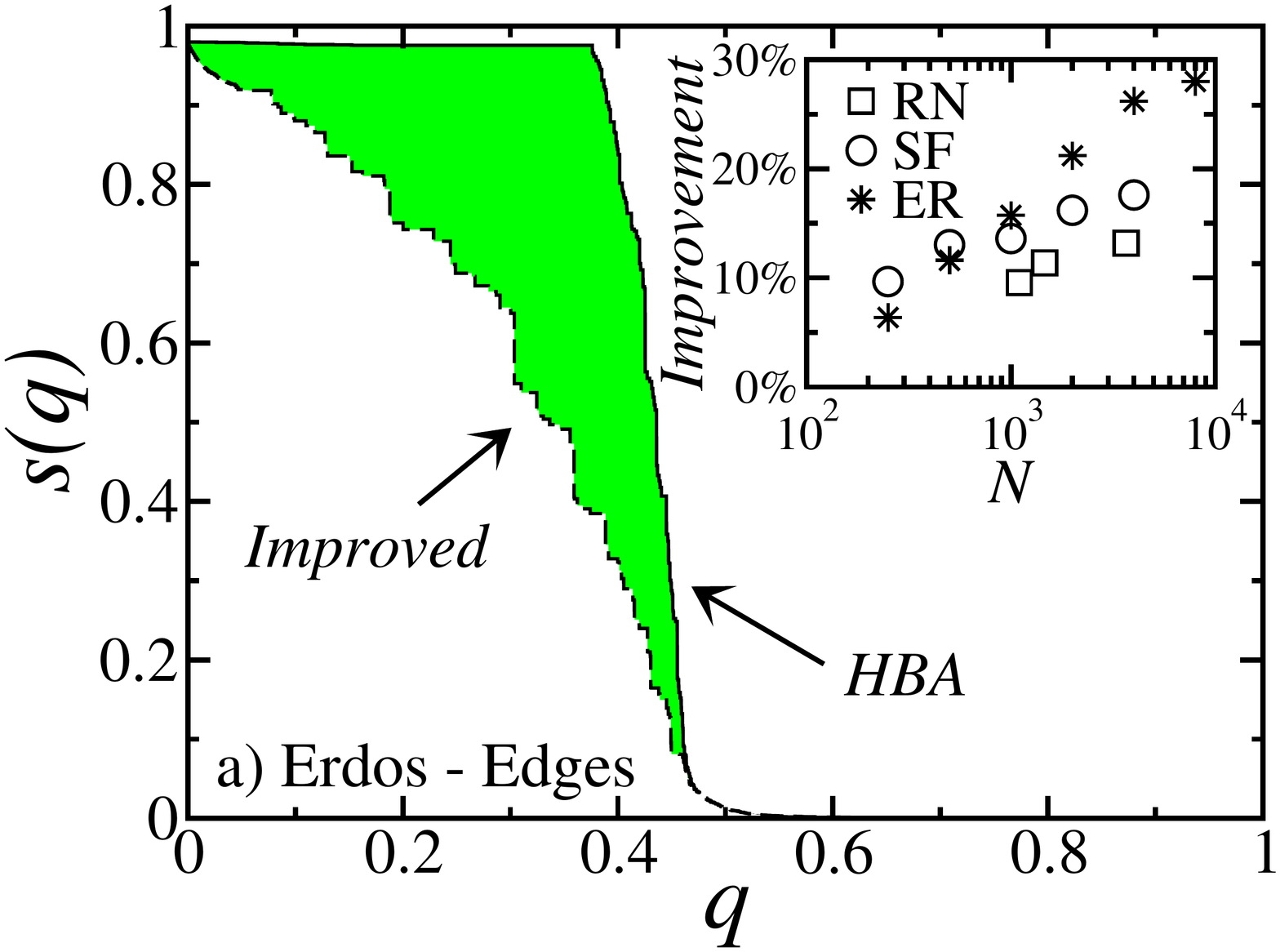} \hspace{1cm}
  \includegraphics[width=4.75cm,angle = -90]{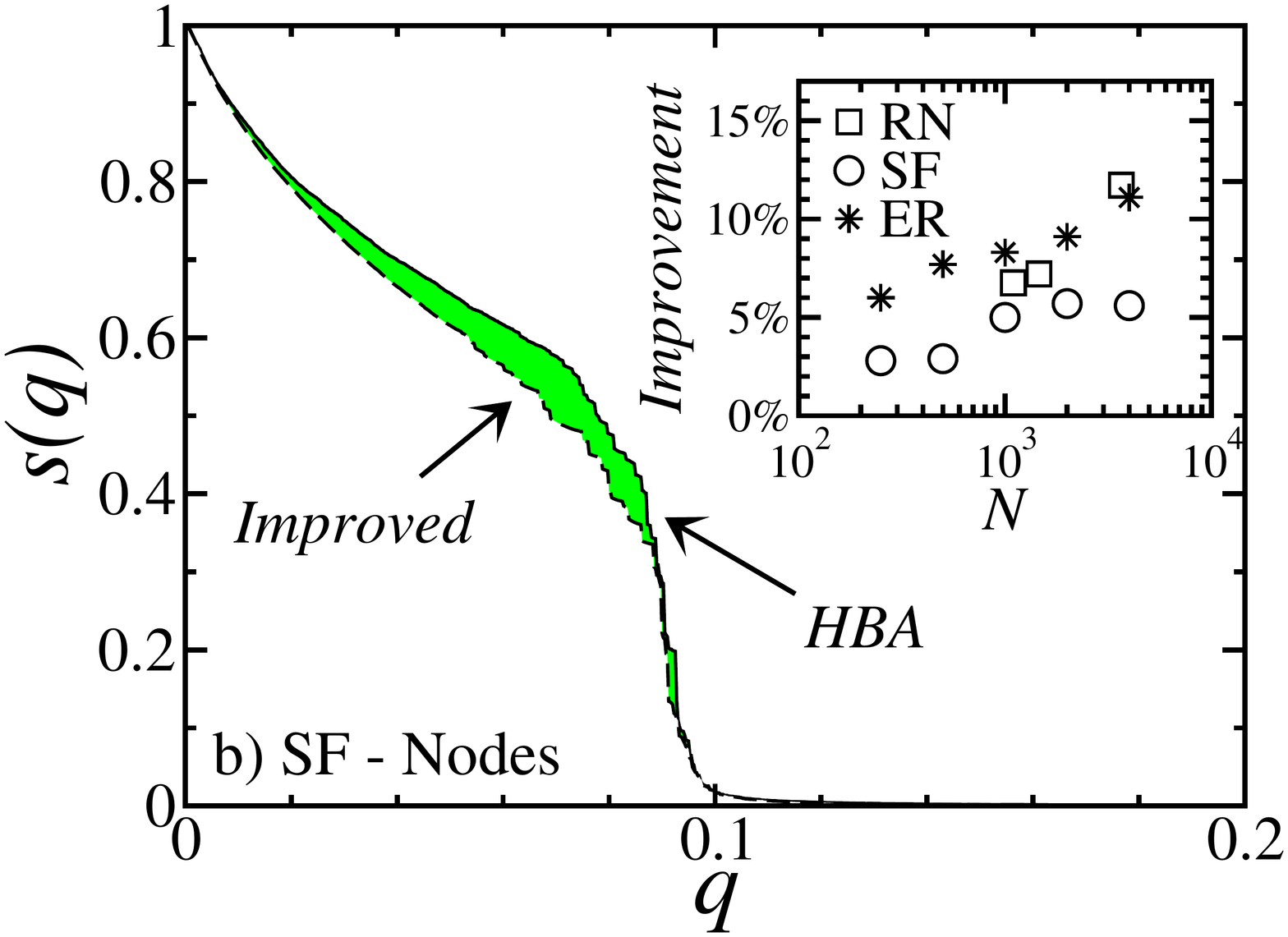}
  \caption{(Color online) Demonstration that betweenness-based immunization can be significantly improved for (a) Erd\H{o}s-R\'enyi (ER) network with $N = 8000$ and $M = 16 000$ where the links are immunized and (b) a scale-free (SF) network with $N = 4000$ and $\gamma = 2.5$ where the nodes are immunized. We plot the fraction of the largest connected cluster of non immunized nodes (which potentially can be infected), $s(q) = S(q)/N$, vs the fraction of the immunized nodes or edges, $q$, for both immunization strategies according to their betweenness (full lines) or according to our improved immunization sequence (dashed lines). Note that the area below the curves represent our measure for susceptible size, $R$ \lbrack Eq. (1)\rbrack. The network size dependences are shown in the insets as well as the results for the three real networks (RN). Our results show that the bigger the network the larger the improvement of our approach.}
\label{fig:2}
\end{figure*}
To verify that the improvements are not only finite size effects, we study different model networks with different numbers of nodes and edges. In Fig. \ref{fig:2}(a), the efficiency of our method is shown on the example of Erd\H{o}s-R\'enyi networks with $N = 8000$ nodes and $M = 16 000$ edges ($\langle k \rangle = 4$). The full line is the fraction of nodes in the largest connected cluster after removing a fraction $q$ of edges, according to the high betweenness immunization strategy and the dashed line corresponds to our improved immunization strategy. Not only is the overall improvement of $R$ about $30\%$ (see inset of Fig. \ref{fig:2}(a)), but also the largest component that can be infected is reduced by up to a factor of $5$ compared to high-betweenness immunization for the practical case of a small number of available immunization doses. {\color{black}In the inset, the size dependence of the improvement of edge immunization is shown for the three real networks, Erd\H{o}s-R\'enyi networks, and scale-free networks with $\gamma = 2.5$. The improvement is calculated by the relative reduction of the susceptible size.} Interestingly, the larger the system, the better the improvement.\\
In Fig. \ref{fig:2}(b) we demonstrate the efficiency of our method on a scale-free network with $N = 4000$ nodes and an exponent $\gamma = 2.5$. Here we immunize the nodes of the network instead of the edges among them. Although, in general, the efficiency of our method is higher in the case of edge immunization, the overall improvement of $R$, in this case, is about $6\%$. The size dependence, shown in the inset, is similar to that for edge immunization for both model and real networks. We have also investigated the effect of the density of edges (degree) in Erd\H{o}s-R\'enyi networks on the improvement and found that it has no major impact on the efficiency of our strategy.\\
\section{Conclusions}
In summary, we introduced a different approach for determining an efficient immunization strategy. We showed that our method is significantly more efficient in preventing disease spreading compared to the high-betweenness method, which was so far believed to be the most efficient. Our method outperforms other immunization strategies for both node and link immunization. We showed this by studying three different performance measures, the average infection probability of SIR model diseases, the susceptible size, and the relative number of saved immunization doses. All of these three measurements indicate that the immunization significantly improves with our strategy for all networks studied. Such improvement could result in the saving of many human lives and resources. For the most important network responsible for global disease spreading, the global airline network, we showed that with our approximated approach the disease spread may be reduced significantly with a relatively small effort. Moreover, we found that the efficiency of our strategy increases with system size.
\section{Acknowledgments}
We acknowledge financial support from the ETH Competence Center 'Coping with Crises in Complex Socio-Economic Systems' (CCSS) through ETH Research Grant CH1-01-08-2 and the Swiss National Science Foundation under contract 200021 126853. S.H. acknowledges support from the Israel Science Foundation, DTRA, DFG, ONR and the Epiwork EU project.\\
This research uses data from Add Health, a program project designed by J. Richard Udry, Peter S. Bearman, and Kathleen Mullan Harris, and funded by a grant P01-HD31921 from the National Institute of Child Health and Human Development, with cooperative funding from 17 other agencies. Special acknowledgment is due Ronald R. Rindfuss and Barbara Entwisle for assistance in the original design. Persons interested in obtaining data files from Add Health should contact Add Health, Carolina Population Center, 123 W. Franklin Street, Chapel Hill, NC 27516-2524 (addhealth@unc.edu).

\end{document}